# In Vivo GABA Detection by Single Pulse Editing with One Shot

Li An, Sungtak Hong, Maria Ferraris Araneta, Tara Turon, Christopher S. Johnson, and Jun Shen

Molecular Imaging Branch, National Institute of Mental Health,

National Institutes of Health, Bethesda, MD

Corresponding author:

Li An

Building 10, Room 3D46

10 Center Drive, MSC 1216

Bethesda, MD 20892-1216

Phone: (301) 594-6868

Email: li.an@nih.gov

# Abstract

Over the past two decades, magnetic resonance spectroscopy with two-shot difference editing has been widely employed to characterize altered levels of GABA, the primary inhibitory neurotransmitter in the brain, in various neuropsychiatric disorders. This conventional technique detects the GABA H4 resonance, making it unsuitable for investigating GABA metabolism. It also suffers from subtraction artifacts, signal loss, and significant contamination by macromolecules. Here, we introduce a single-shot method for detecting GABA H2, effectively overcoming these difficulties. Since GABA turnover initiates at its protonated C2 and unprotonated C1 positions, we demonstrate, for the first time, noninvasive real-time monitoring of GABA metabolism in the human brain, utilizing GABA H2 as a highly sensitive reporter for GABA C2. This new method not only enhances the quantitative measurement of GABA levels but also opens up a new avenue to probe the metabolic processes underlying alterations in GABA levels in patients.

# Main

γ-aminobutyric acid (GABA) is the major inhibitory neurotransmitter in the brain and plays a key role in the excitation-inhibition balance[1]. In the central nervous system (CNS), significant variations in GABA concentration exist accompanied by its highly active metabolism. Altered levels of GABA have been indicated in many neuropsychiatric disorders, including epilepsy[2], schizophrenia[3-5], bipolar disorder[6], and major depression[7-9]. However, direct detection of GABA in vivo using proton magnetic resonance spectroscopy (MRS) without spectral editing is difficult due to significant spectral overlap, which leads to detrimental correlations between overlapping signals[10].

J-difference editing[2] techniques have been widely used for in vivo GABA detection in the past two decades[11-23]. These techniques selectively invert the GABA H3 spins at 1.89 ppm at an optimal echo time (TE) of ~68 ms on alternate acquisitions, revealing its H4 signal at 3.01 ppm through subtraction. However, difference editing of GABA is prone to creatine (Cr) subtraction artifacts caused by head movement and system instability[13, 24]. It also leads to significant signal loss because the central peak of the GABA H4 triplet is largely cancelled in the difference spectrum[14]. Additionally, the GABA H4 signal is heavily contaminated by the co-edited macromolecule signal at 3.0 ppm[25]. Although efforts have been made to eliminate this macromolecule contamination[12], it comes at the expense of considerably degraded precision for GABA quantification[20].

Several single-shot techniques, such as double quantum filtering methods, have also been devised for in vivo GABA detection[26-32]. These single-shot techniques are less vulnerable to head movement and system instability. However, similar to J-difference editing, they also result in considerable signal loss. Furthermore, since double quantum filtering eliminates all singlets, spectral phasing of the filtered spectra may become difficult, leading to potential quantification errors or significantly increased technical complexity[30].

Although proton MRS has been widely used to characterize altered GABA levels in various brain disorders, the metabolic processes underlying GABA level alterations remain beyond the reach of current clinical MRS technologies. A major challenge in measuring GABA metabolism is that the conventional GABA editing methods rely on detecting GABA H4, whereas GABA turnover starts at its protonated C2 and unprotonated C1[33]. Since gaining an understanding of the causal mechanisms underlying GABAergic abnormalities is expected to provide crucial insights into the biochemical basis of many neuropsychiatric disorders and potentially guide the development of treatment strategies, there is a pressing need for a new MRS strategy to detect GABA H2, which can serve as a high-sensitivity reporter of GABA C2 for monitoring GABA metabolism.

We introduce a novel editing method, termed Single Pulse Editing with One Shot (SPEOS), for in vivo detection of GABA at 7 Tesla. A single always-on 180° editing pulse is applied at 1.89 ppm to refocus the GABA H2 and H4 signals, with the GABA H2 signal at 2.28 ppm fully preserved and serving as the primary target for GABA quantification. The glutamate (Glu) H4 signal at 2.34 ppm partially overlaps with the GABA H2 signal. Spectral overlap generally interferes with the detection of overlapping signals, leading to unwanted spectral correlations[10, 34]. We show that, through pulse sequence design, the generally much stronger Glu H4 signal can be reduced to a level comparable to the GABA H2 signal. Additionally, the correlation between Glu and GABA, arising from spectral overlap, can be practically eliminated by orthogonalizing the lineshapes of Glu and GABA. This minimizes the interference of the Glu H4 signal with the detection of the GABA H2 signal. Furthermore, metabolite nulling experiments were conducted to confirm the minimized contribution from macromolecules to the observed GABA H2 signal. Using SPEOS, we demonstrate, for the first time, noninvasive real-time monitoring of GABA metabolism in the human brain.

## Results

To minimize the GABA-Glu correlation while maximizing the target GABA H2 signal, a single chemical shift-selective editing pulse is used (vide infra). However, the use of a single editing pulse causes a nonlinear phase distortion of the MRS spectra. The real and imaginary parts of the numerically calculated Bloch-Siegert phasor, as functions of chemical shift, are plotted in Supplementary Fig. S1a. The Bloch-Siegert phase shift (in degrees), as a function of chemical shift, is plotted in Supplementary Fig. S1b. Supplementary Fig. S1 shows that the Bloch-Siegert phase shift is substantial near the editing frequency (1.89 ppm) but gradually diminishes for signals resonating further away from the editing frequency.

Here, we demonstrate the accurate correction of the Bloch-Siegert phase shift, eliminating the necessity for a second editing pulse. This paved the way for minimizing the spectral overlapping effect of Glu H4, ensuring optimal detection of GABA H2. The numerically calculated spectra without (a) and with (b) Bloch-Siegert phase shift correction are shown in Supplementary Fig. S2, where the GABA spectra and the spectra of the sum of N-acetylaspartate (NAA), GABA, Glu, Cr, and choline (Cho) are plotted. Without Bloch-Siegert phase shift correction (a), the NAA singlet and the GABA H2 are significantly affected by the Bloch-Siegert phase shift. However, the effect of the phase shift is much smaller for other resonance signals further away from 1.89 ppm. The Cr, Cho, and GABA H4 peaks show minimal influence from the Bloch-Siegert phase shift. With Bloch-Siegert phase shift correction (b), the phase shifts for the NAA singlet and all other resonance signals including GABA H2 are successfully removed.

To minimize the interference from the Glu signal with GABA detection, we sought to both minimize the signal intensity of the interfering Glu H4 peak and orthogonalize the Glu H4 and GABA H2 lineshapes (see GABA-Glu Correlation in Supplementary Information). Density matrix simulated spectra of Glu, GABA, and their sum for six different TE values are displayed in Fig. 1. In the TE range of 68 – 88 ms, the generally much higher Glu H4 peak is reduced to a level comparable to that of the GABA H2 peak. The Glu peak is the smallest at TE = 68 ms and gradually increases when TE becomes longer. The GABA-Glu correlation coefficient reaches approximately zero (Pearson's correlation coefficient r = 0.04) at TE = 76 ms. Supplementary Fig. S3 shows the simulated spectra of Glu, GABA, and their sum at TE = 76 ms with 7 Hz and 11 Hz line broadening. The GABA-Glu correlation coefficient remains minimal for both 7 Hz and 11 Hz line broadening. At a longer TE (88 ms), the GABA-Glu correction coefficient is similarly minimized (see Fig. 1). However, the peak amplitude for GABA H2 at TE = 88 ms is 9.5% lower than at TE = 76 ms. There is also a significantly greater signal loss at TE = 88 ms when $T_2$ relaxation is considered. Therefore, 76 ms was chosen as the optimal TE for detecting GABA H2.

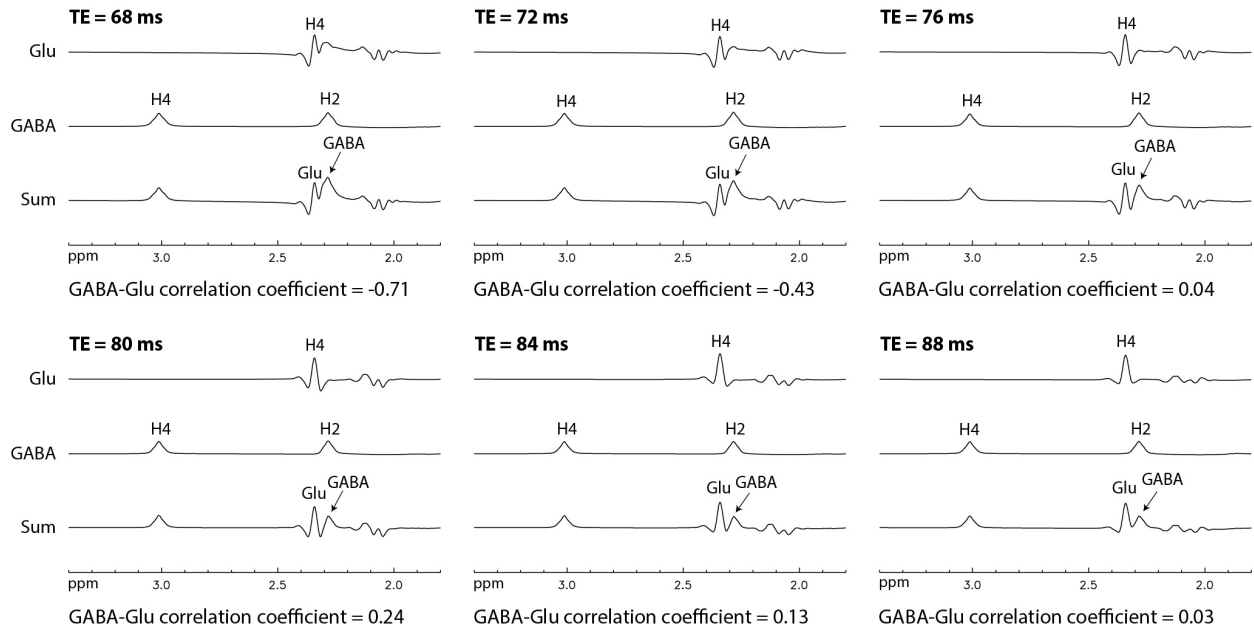

**Fig. 1:** Numerically calculated spectra of Glu, GABA, and their sum for six different TE values. The relative concentrations of Glu and GABA were 13 and 1, respectively. $T_2$ relaxation was ignored, and all spectra were line broadened by 9 Hz using the Lorentzian lineshape.

In vivo test and re-test measurements were conducted at 7 Tesla using a $2 \times 3.5 \times 2$ cm$^3$ voxel in the anterior cingulate cortex (ACC) of six healthy participants. The results are displayed in Fig. 2. As expected, the Glu H4 peak is reduced to a level comparable to that of GABA H2. The test and re-test spectra are highly consistent, and the GABA H2 peak is prominent in all spectra. The linewidth of the total creatine (tCr) singlet is found to be $10.3 \pm 0.8$ Hz, falling within the 7 – 11 Hz range used in the simulation, as shown in Supplementary Fig. S3.

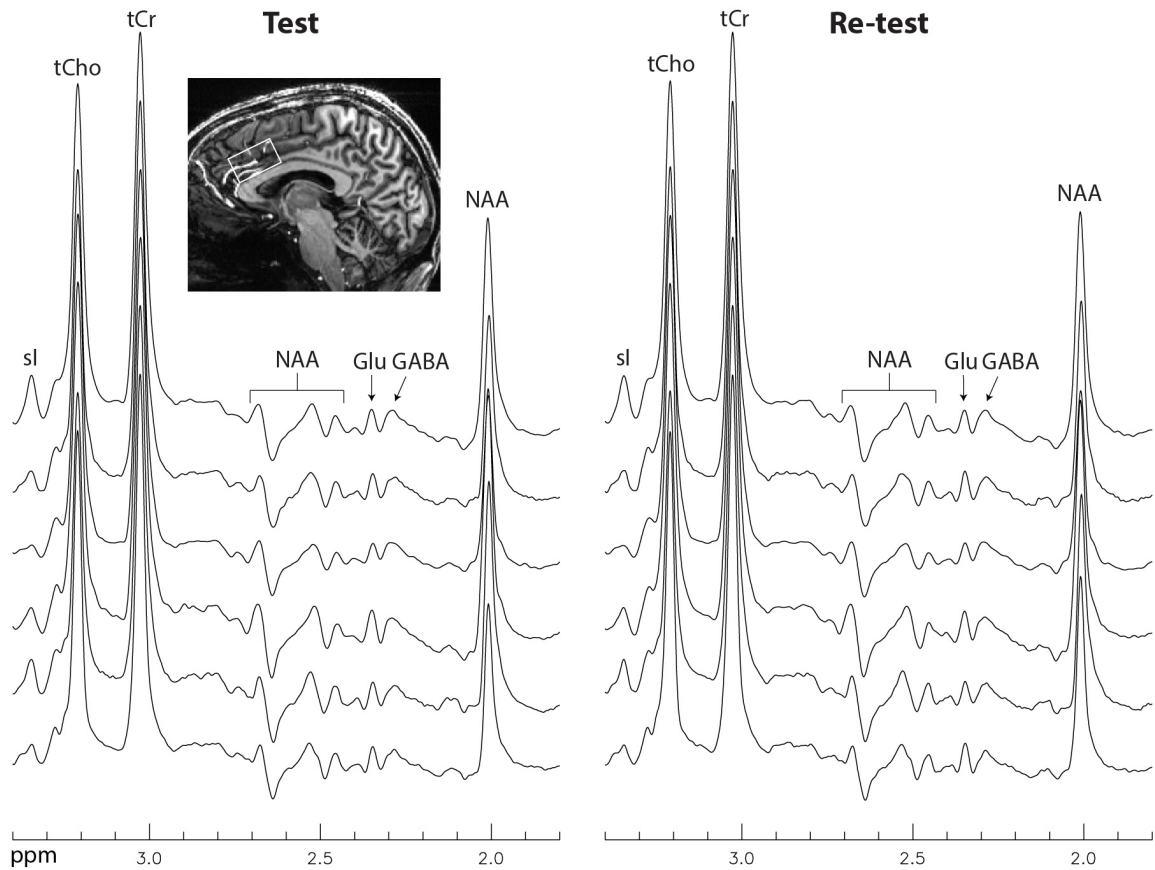

**Fig. 2:** In vivo spectra acquired from the ACC of six healthy participants using the SPEOS pulse sequence (Supplementary Fig. S7). Spectra for the test and retest measurements of each participant are displayed on the left and right, respectively. No linebroadening was applied. Voxel size = 2 × 3.5 × 2 cm$^3$; Repetition time (TR) = 2.5 s; TE = 76 ms; spectral width = 4000 Hz; number of data points = 1024; number of averages = 116; and total scan time = 5 min per spectrum. tCho: total choline; sI: scyllo-inositol.

We conducted inversion-recovery metabolite-nulling experiments on three participants to compare macromolecules at GABA H4 and H2 resonances. Fig. 3 displays the in vivo spectra without inversion recovery, the metabolite-nulled spectra, and the macromolecule spectra acquired from the three participants. In both the metabolite-nulled spectra (b) and the macromolecule spectra (c), the M6 peak[25] at ~2.3 ppm, which overlaps with the GABA H2 peak, is much weaker than the M7 peak at ~3.0 ppm. The latter overlaps with the GABA H4 peak in conventional J-difference GABA editing. Therefore, SPEOS enables GABA quantification with greatly reduced macromolecule contamination, in contrast to the two-shot difference editing techniques that measure GABA H4.

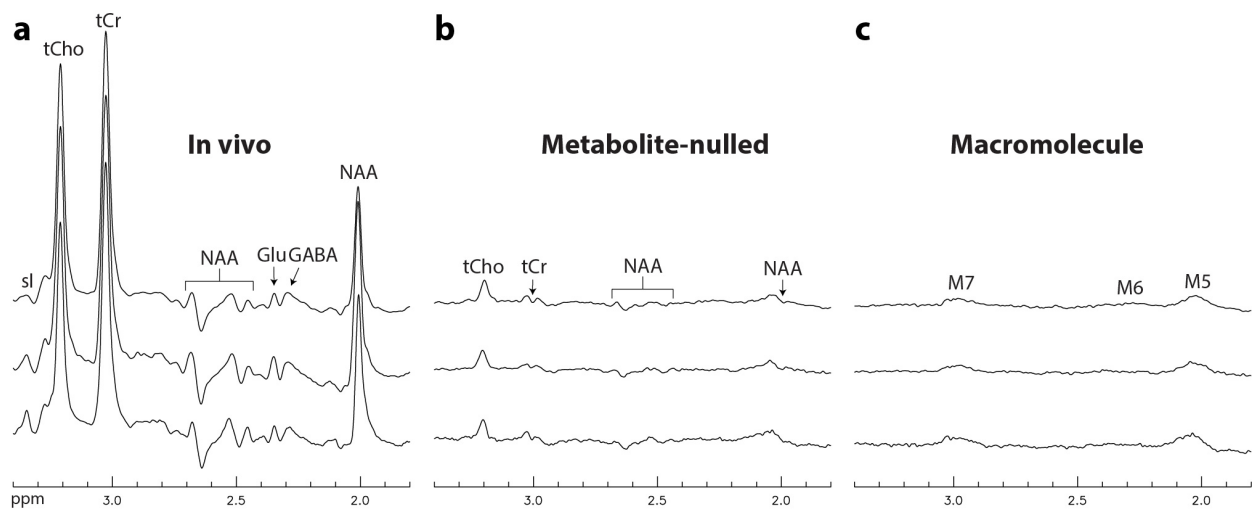

**Fig. 3:** In vivo spectra and corresponding metabolite-nulled spectra acquired with inversion recovery from three healthy participants. (**a**) In vivo spectra acquired without inversion recovery. (**b**) Metabolite-nulled spectra acquired with inversion recovery. TR = 3 s; inversion time = 850 ms; TE = 76 ms; number of averages = 96; and total scan time = 5 min. (**c**) Macromolecule spectra obtained by subtracting the fitted residual metabolite signals from the metabolite-nulled spectra.

In vivo spectra and fitting details for a healthy participant are provided in Supplementary Fig. S4. The in vivo test and re-test spectra, along with their corresponding fitted spectra, demonstrate a high level of consistency. Table 1 lists the in vivo metabolite ratios (/[tCr]) of GABA, Glu, tCr, and tCho computed using a total of 12 MRS measurements from the six healthy participants. The Cramer-Rao lower bounds (CRLBs) and within-subject coefficients of variation (CVs) are found to be quite low for the 5 min scan time and the 14 mL voxel placed in the ACC. The high consistency of the test and retest results, as well as the very low CVs and CRLBs, attests to the robustness of our single-shot approach, which does not rely on data subtraction for GABA detection. These characteristics of SPEOS are crucial for monitoring GABA metabolism, which generally involves much longer scans and therefore demands high immunity to patient movement and system instability.

To demonstrate SPEOS's capability for noninvasive real-time monitoring of GABA metabolism in the human brain, we conducted scans on healthy participants both before and after the administration of $^{13}$C-labeled glucose. The time-course $^1$H spectra acquired from a participant before and after oral intake of [U-$^{13}$C]glucose are presented in Fig. 4. An additional example of time-course spectra, showing the $^{13}$C labeling of GABA, is provided in Supplementary Fig. S5. These results clearly indicate a decreasing GABA H2 peak in the time-course $^1$H spectra, reflecting the metabolizing GABA molecules as $^{12}$C at the GABA C2 position being replaced by $^{13}$C originating from the $^{13}$C-labeled glucose.

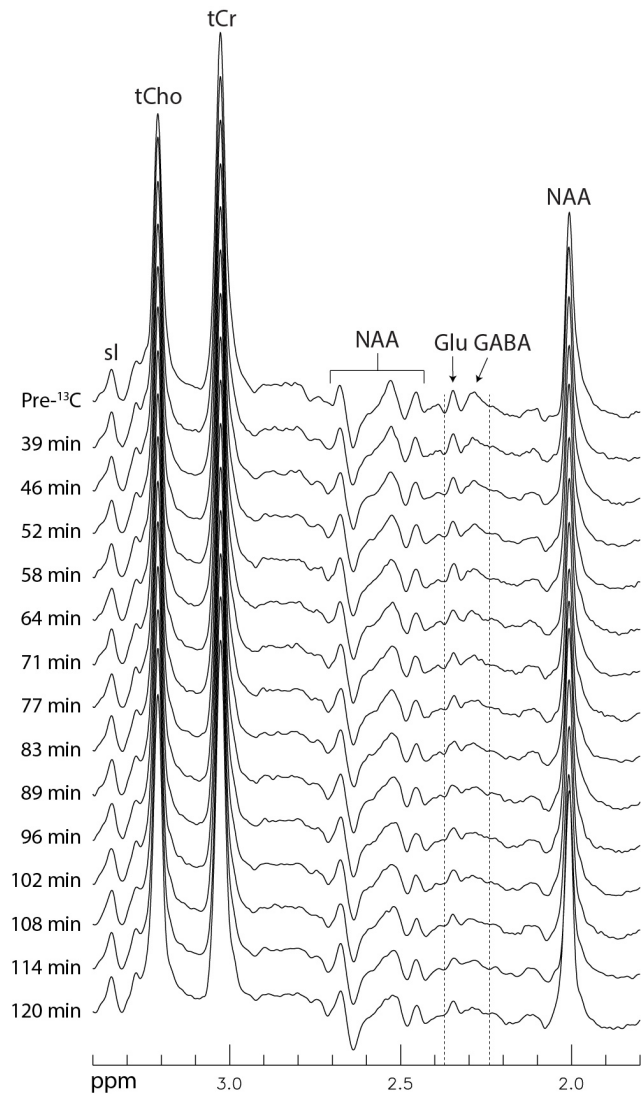

**Fig. 4:** Time-course $^1$H spectra acquired from the ACC of a healthy participant following oral administration of [U-$^{13}$C]glucose. The pair of dashed lines indicate the spectral range where Glu H4 and GABA H2 resonate, whose peak amplitudes decrease as the corresponding $^{12}$Cs being replaced by $^{13}$Cs. The time-course acquisition was initiated after the participant had drunk the glucose solution, rested, and then re-entered the magnet. Each spectrum was acquired using the same pulse sequence parameters as given in Fig. 2.

## Discussion

Existing J-difference editing techniques typically employ a pair of chemical shift-selective editing pulses. The second editing pulse refocuses the Bloch-Siegert phase shift induced by the first editing pulse. In contrast, the SPEOS technique utilizes only a single editing pulse, positioned between the two slice-selective refocusing pulses. Consequently, the reconstructed spectrum exhibits a Bloch-Siegert phase shift as a function of chemical shift, causing nonlinear spectral distortions. To correct for this Bloch-Siegert phase shift, the complex-conjugate of the Bloch-Siegert phasor function was computed using density matrix simulation[35] and multiplied to the spectrum. For our technique, the use of a single editing pulse offers several benefits, including a shorter minimum TE and greater flexibility in positioning the editing pulse. These advantages are crucial for optimizing the pulse sequence timing to minimize the GABA-Glu correlation while maintaining a near-maximized GABA H2 signal.

By toggling the single editing pulse on and off for alternate acquisitions, it is possible to convert the proposed sequence into a J-difference editing sequence that uses a single editing pulse and two shots for measuring GABA H4. Numerically calculated spectra for a J-difference GABA editing pulse sequence with a single editing pulse are shown in Supplementary Fig. S6. The difference spectrum for the sum of metabolites reveals the GABA H4 peak at 3.01 ppm with a negligible residual Cr signal at 3.03 ppm in the absence of patient movement or system instability. This also shows that the use of a pair of editing pulses is not necessary for J-difference editing of GABA H4 at 7 Tesla.

Due to both the strong scalar coupling between the two geminal H4 protons of Glu and the weak couplings between the H3 and H4 protons of Glu[36], significant changes in the spectral pattern of Glu H4 occur for different TE values across the 68 - 88 ms TE range (Fig. 1). At TE = 68 ms, a substantial positive peak appears at the upfield end of the Glu H4 signal. This positive peak dominates the overlap between the Glu H4 signal and the positive GABA H2 signal, resulting in a negative correlation coefficient with a large magnitude (r = -0.71). This correlation arises because an overdetermination (or underdetermination) of GABA H2 is compensated by an underdetermination (or overdetermination) of Glu H4, and vice versa. As TE increases, this positive upfield peak decreases while a negative upfield peak of Glu H4 becomes more significant. This change in the spectral pattern of Glu H4 reduces the magnitude of the correlation coefficient as contributions to the GABA-Glu correlation from the positive and negative upfield peaks of Glu H4 offset each other. At TE = 76 ms, the positive and negative contributions of Glu H4 to the Glu-GABA correlation cancel each other out because the basis spectra of Glu and GABA become nearly orthogonal ($\mathbf{B_1} \cdot \mathbf{B_2} \approx 0$, see GABA-Glu Correlation in the Supplementary Information), minimizing spectral interference from Glu H4 with the detection of GABA H2 (r = 0.04). When TE reaches 88 ms, the correlation coefficient approaches zero again. However, longer TEs were not used to avoid GABA signal loss due to J-evolution and greater $T_2$ relaxation effect at longer TEs.

It is well known that the GABA H4 signal at 3.01 ppm detected by J-difference editing has a significant contribution from the macromolecule resonance at 3.0 ppm, which has a J-coupling partner at 1.7 ppm[12, 14, 16, 17, 20, 25]. As shown in Fig. 3, the M6 signal at 2.3 ppm exhibits a much lower and flatter profile than the M7 peak at 3.0 ppm. Consequently, macromolecule contamination of the GABA H2 peak at 2.28 ppm is greatly reduced compared to that of the GABA H4 peak at 3.01 ppm. The GABA/tCr ratio obtained from this study is 0.07 ± 0.01, which is 37% lower than 0.116 ± 0.014 obtained from a multi-institutional study of GABA using difference editing of GABA H4 at 3 Tesla[20]. This is consistent with significantly reduced macromolecule contribution using our method.

GABA metabolism plays a key role in many neuropsychiatric disorders, such as epilepsy and major depressive disorder[37, 38]. Clinical studies have demonstrated that vigabatrin, a highly specific GABA-transaminase inhibitor, is an effective anticonvulsant drug for drug-resistant partial epilepsy, with improved seizure control correlating with the altered GABA catabolism and elevated GABA concentration in the brain[39-41]. Although reduced prefrontal GABA levels are found in patients with major depression, GABA levels are normal in remitted patients[7, 42-46], implying a critical role of GABA metabolism in symptom improvement in major depressive disorder. These examples of GABA level alterations associated with disease states and treatment provide a compelling rationale for investigating the metabolic processes underlying altered GABA levels. Our results, as shown in Figs. 4 and S5, demonstrate that GABA metabolism can be monitored in real time by SPEOS with the high sensitivity and high spatial resolution of proton MRS while using commercially available hardware. This new MRS capability, as described in this work, is expected to provide a unique noninvasive tool for assessing metabolic fluxes underlying altered GABA levels, thereby aiding the development of treatments targeting the GABAergic system.

In summary, we have developed SPEOS, a novel GABA editing method that uses single pulse editing with one shot for in vivo detection of the full GABA H2 signal at 7 Tesla. The pulse sequence timing has been crafted to minimize interference from Glu H4 with the detection of GABA H2. This new GABA editing method overcomes difficulties such as the loss of GABA signal caused by the cancellation of the middle peak of the GABA H4 triplet, as well as subtraction artifacts commonly associated with difference editing techniques. Additionally, it effectively minimizes contamination of the GABA signal by macromolecules due to the significantly weaker macromolecule signal at 2.3 ppm. More importantly, it enables noninvasive, real-time monitoring of GABA metabolism in the human brain, opening up a new avenue for clinical investigation of the metabolic mechanisms underlying altered GABA levels in neuropsychiatric disorders.

## Methods

### Pulse Sequence

A schematic diagram of the SPEOS pulse sequence is displayed in Supplementary Fig. S7. This pulse sequence was created by incorporating an 180° editing pulse between the two 180° slice-selective refocusing pulses of a point-resolved spectroscopy (PRESS) sequence[47]. The editing pulse has a duration of 15 ms, and its amplitude profile was generated by truncating a Gaussian function at one standard deviation on each side, resulting in a full width half maximum (FWHM) bandwidth of 73 Hz. The detailed timing diagram of the pulse sequence and the frequency profile of the editing pulse are shown in Supplementary Figs. S8 and S9, respectively[48].

### Bloch-Siegert Phase Shift Correction

The single editing pulse induces a Bloch-Siegert phase shift[49] as a function of chemical shift in the reconstructed spectrum, which was numerically computed using density matrix simulation[35]. Because the Bloch-Siegert phase shift is solely determined by the editing pulse and independent of the slice-selective excitation and refocusing pulses in the pulse sequence (Supplementary Fig. S7), these three slice-selective pulses were replaced with ideal pulses, with the associated slice-selection and crusher gradients removed, to drastically speed up the simulation. The 15 ms long experimental editing pulse, along with a pair of crusher gradients with opposite polarization, were used in the simulation. For the pair of crusher gradients, the simulation used 2000 spatial points[35]. The spectrum of a single spin was simulated at 6087 different chemical shift positions in the range of 0 – 5 ppm with an increment of $8.21\times10^{-4}$ ppm (0.244 Hz). For each of the 6087 chemical shift positions, the phasor of the single peak in the reconstructed spectrum was determined by fitting a Lorentzian curve to the peak. This numerically calculated phasor function was used to correct for the Bloch-Siegert phase shift. Specifically, the spectrum was multiplied by the complex-conjugate of the phasor function to remove the Bloch-Siegert phase shift.

To demonstrate the effect of Bloch-Siergert phase shift correction, density matrix simulation with high spatial digitization[35, 50] was employed to calculate the spectra of NAA, GABA, Glu, Cr, and Cho. An ideal pulse was used for the excitation pulse without any slice-selection gradient to speed up the simulation without significantly affecting the simulated spectra. The slice-selective refocusing pulses and the single chemical shift-selective editing pulse employed in the in vivo experiments were used in the simulation. The corresponding slice-selection and crusher gradients were digitized with $200 \times 200 \times 2000$ spatial points[35]. After obtaining the metabolite spectra from the density matrix simulation, the Bloch-Siegert phase shift correction was applied to the metabolite spectra to correct the phase shift (see Supplementary Fig. S2).

**Pulse Sequence Timing Optimization**

To minimize the correlation between GABA and Glu, density matrix simulation with high spatial digitization was utilized to calculate the basis spectra of Glu and GABA at a series of different TE, $TE_1$, and $T_d$ values (see Figure S7). The Supplementary Equation (S7) shows that the Pearson's correlation coefficient between GABA H2 and Glu H4, arising from spectral overlap, is the opposite value of the normalized dot product of $B_1$ and $B_2$. Here, the vectors $B_1$ and $B_2$ contain the basis spectra of GABA and Glu, respectively, in the range of 2.21 – 2.36 ppm. To determine the optimal TE for detecting GABA with minimized interference from Glu H4, the TE of the single-shot GABA editing pulse sequence was varied from 68 to 88 ms with an increment of 1 ms. The correlation coefficient between the GABA and Glu basis spectra was calculated for each TE. Additionally, $TE_1$ and $T_d$ were adjusted to minimize GABA-Glu correlation for each TE. The TE value that resulted in the overall minimum GABA-Glu correlation was selected as the optimal TE for detecting GABA H2.

**In Vivo Experiments**

Six healthy participants (two females and four males; age = 39 ± 11 years) were recruited for the study using a Siemens Magnetom 7 Tesla scanner. Written informed consent was obtained from all participants before the study following the procedures approved by the Institutional Review Board (IRB) of the National Institute of Mental Health (NIMH; NCT01266577; NCT00109174). All experimental protocols and methods were performed in accordance with the guidelines and regulations of the NIH MRI Research Facility. A three-dimensional $T_1$-weighted magnetization prepared rapid gradient echo (MPRAGE) image was acquired with TR = 3 s, TE = 3.9 ms, data matrix = 256 × 256 × 256, and spatial resolution = 1 × 1 × 1 mm³. The 2 × 3.5 × 2 cm³ MRS voxel in the ACC had a water linewidth of 12.0 ± 0.6 Hz. The numerically optimized pulse sequence with TE = 76 ms, $TE_1$ = 59.1 ms, and $T_d$ = 22 ms was used to acquire the MRS data. Seven variable power RF pulses (sinc-Gauss pulse, duration = 26 ms, bandwidth = 105 Hz) with optimized relaxation delays (VAPOR) were used for water suppression. Using a TR of 2.5 s and 116 signal averages, the total time for one MRS scan was 5 min. Test and re-test MRS scans were performed on the same voxel for each participant.

Three of the six healthy participants were also scanned with oral administration of [U-$^{13}$C]glucose. After the pre-$^{13}$C MRS scan was finished, the participants exited the scanner and drank 20% w/w 99% enriched [U-$^{13}$C]glucose solution at a dosage of 0.75 g [U-$^{13}$C]glucose per kg of body weight following procedures described in our previous study of carbonic anhydrase-catalyzed $^{13}$C magnetization transfer and references therein[51]. After a rest period, the participants reentered the scanner and the post-$^{13}$C MRS scans were carried out.

The 32-channel free induction decay (FID) data were combined into single-channel FIDs using the generalized least square (GLS) method[52], in which coil sensitivities were computed from the unsuppressed water signals acquired with two acquisitions. The unsuppressed water signals were also used to correct for the phase errors in the combined FIDs caused by zero-order eddy currents[53]. The frequency deviation for each acquisition was determined and corrected by fitting the magnitude of the tCr and tCho peaks with two Voigt curves. The Bloch-Siegert phase shift in the reconstructed spectrum, which was the average of all acquisitions, was corrected using the numerically calculated Bloch-Siegert phasor function. The reconstructed spectrum was fitted in the range of 1.8 – 3.4 ppm by a linear combination of numerically calculated metabolite basis spectra and a cubic spline baseline. Chemical shifts and coupling constants were obtained from Ref.[14] for GABA, from Ref.[54] for Glu, from Ref.[55] for glutathione (GSH), and from Ref.[56] for acetate (Ace), NAA, N-acetylaspartylglutamate (NAAG), glutamine (Gln), aspartate (Asp), Cr, phosphocreatine, phosphocholine, glycerophosphocholine, taurine (Tau), myo-inositol (mI), and scyllo-inositol (sI). Basis spectra for 31 frequency deviation values ranging from -15 Hz to 15 Hz at 1 Hz intervals were computed. The basis spectra used in the fitting were computed as the weighted average of the basis spectra corresponding to the 31 frequency deviation values, where the experimentally measured frequency deviation histogram was used as the weighting function[35].

The GABA H2 signal (2.21 – 2.36 ppm) was given a weighting factor 10 times that of the unresolved GABA H4 signal (2.94 – 3.08 ppm) in the fitting process to make GABA H2 the dominant signal for GABA quantification.

Metabolite-nulled spectra were also obtained from three out of the six healthy participants. To acquire these spectra, a hyperbolic-secant inversion RF pulse (duration = 9.6 ms, bandwidth = 1.75 kHz, inversion time = 850 ms) was incorporated into the proposed pulse sequence (Supplementary Fig. S7). The residual metabolite signals in the metabolite-nulled spectra were determined by spectral fitting and subtracted from the metabolite-nulled spectra to generate the macromolecule spectra with metabolite signals removed.

## References

1. McCormick, D.A. GABA as an inhibitory neurotransmitter in human cerebral cortex. *J Neurophysiol* **62**, 1018-1027 (1989).
2. Rothman, D.L., Petroff, O.A., Behar, K.L. & Mattson, R.H. Localized 1H NMR measurements of gamma-aminobutyric acid in human brain in vivo. *Proc Natl Acad Sci U S A* **90**, 5662-5666 (1993).
3. Ongur, D., Prescot, A.P., McCarthy, J., Cohen, B.M. & Renshaw, P.F. Elevated gamma-aminobutyric acid levels in chronic schizophrenia. *Biol Psychiatry* **68**, 667-670 (2010).
4. Rowland, L.M. et al. GABA predicts inhibition of frequency-specific oscillations in schizophrenia. *J Neuropsychiatry Clin Neurosci* **25**, 83-87 (2013).
5. Yoon, J.H. et al. GABA Concentration Is Reduced in Visual Cortex in Schizophrenia and Correlates with Orientation-Specific Surround Suppression. *Journal of Neuroscience* **30**, 3777-3781 (2010).
6. Bhagwagar, Z. et al. Reduction in occipital cortex gamma-aminobutyric acid concentrations in medication-free recovered unipolar depressed and bipolar subjects. *Biol Psychiatry* **61**, 806-812 (2007).
7. Hasler, G. et al. Reduced prefrontal glutamate/glutamine and gamma-aminobutyric acid levels in major depression determined using proton magnetic resonance spectroscopy. *Arch Gen Psychiatry* **64**, 193-200 (2007).
8. Bhagwagar, Z. et al. Low GABA concentrations in occipital cortex and anterior cingulate cortex in medication-free, recovered depressed patients. *Int J Neuropsychopharmacol* **11**, 255-260 (2008).

9. Sanacora, G. et al. Reduced cortical gamma-aminobutyric acid levels in depressed patients determined by proton magnetic resonance spectroscopy. *Arch Gen Psychiatry* **56**, 1043-1047 (1999).
10. Hong, S. & Shen, J. Neurochemical correlations in short echo time proton magnetic resonance spectroscopy. *NMR Biomed* **36**, e4910 (2023).
11. Keltner, J.R. et al. A technique for detecting GABA in the human brain with PRESS localization and optimized refocusing spectral editing radiofrequency pulses. *Magn Reson Med* **36**, 458-461 (1996).
12. Henry, P.G., Dautry, C., Hantraye, P. & Bloch, G. Brain GABA editing without macromolecule contamination. *Magn Reson Med* **45**, 517-520 (2001).
13. Waddell, K.W., Avison, M.J., Joers, J.M. & Gore, J.C. A practical guide to robust detection of GABA in human brain by J-difference spectroscopy at 3 T using a standard volume coil. *Magn Reson Imaging* **25**, 1032-1038 (2007).
14. Kaiser, L.G., Young, K., Off, D.J.M., Mueller, S.G. & Matson, G.B. A detailed analysis of localized J-difference GABA editing: theoretical and experimental study at 4T. *Nmr in Biomedicine* **21**, 22-32 (2008).
15. Geramita, M. et al. Reproducibility of prefrontal gamma-aminobutyric acid measurements with J-edited spectroscopy. *Nmr in Biomedicine* **24**, 1089-1098 (2011).
16. Andreychenko, A., Boer, V.O., de Castro, C.S.A., Luijten, P.R. & Klomp, D.W.J. Efficient spectral editing at 7 T: GABA detection with MEGA-sLASER. *Magnetic Resonance in Medicine* **68**, 1018-1025 (2012).
17. Edden, R.A., Puts, N.A. & Barker, P.B. Macromolecule-suppressed GABA-edited magnetic resonance spectroscopy at 3T. *Magn Reson Med* **68**, 657-661 (2012).
18. Puts, N.A.J. & Edden, R.A.E. In vivo magnetic resonance spectroscopy of GABA: A methodological review. *Progress in Nuclear Magnetic Resonance Spectroscopy* **60**, 29-41 (2012).
19. Harris, A.D., Puts, N.A., Barker, P.B. & Edden, R.A. Spectral-editing measurements of GABA in the human brain with and without macromolecule suppression. *Magn Reson Med* **74**, 1523-1529 (2015).
20. Mikkelsen, M. et al. Big GABA: Edited MR spectroscopy at 24 research sites. *Neuroimage* **159**, 32-45 (2017).
21. Saleh, M.G. et al. Simultaneous editing of GABA and glutathione at 7T using semi-LASER localization. *Magn Reson Med* **80**, 474-479 (2018).
22. Chan, K.L., Oeltzschner, G., Saleh, M.G., Edden, R.A.E. & Barker, P.B. Simultaneous editing of GABA and GSH with Hadamard-encoded MR spectroscopic imaging. *Magn Reson Med* **82**, 21-32 (2019).
23. Mikkelsen, M. et al. Big GABA II: Water-referenced edited MR spectroscopy at 25 research sites. *Neuroimage* **191**, 537-548 (2019).
24. Evans, C.J. et al. Subtraction artifacts and frequency (mis-)alignment in J-difference GABA editing. *J Magn Reson Imaging* **38**, 970-975 (2013).
25. Behar, K.L., Rothman, D.L., Spencer, D.D. & Petroff, O.A.C. Analysis of Macromolecule Resonances in H-1-Nmr Spectra of Human Brain. *Magnetic Resonance in Medicine* **32**, 294-302 (1994).
26. Wilman, A.H. & Allen, P.S. Yield enhancement of a double-quantum filter sequence designed for the edited detection of GABA. *J Magn Reson B* **109**, 169-174 (1995).

27. Keltner, J.R., Wald, L.L., Frederick, B.D. & Renshaw, P.F. In vivo detection of GABA in human brain using a localized double-quantum filter technique. *Magn Reson Med* **37**, 366-371 (1997).
28. Shen, J., Shungu, D.C. & Rothman, D.L. In vivo chemical shift imaging of gamma-aminobutyric acid in the human brain. *Magn Reson Med* **41**, 35-42 (1999).
29. McLean, M.A. et al. In vivo GABA+ measurement at 1.5T using a PRESS-localized double quantum filter. *Magn Reson Med* **48**, 233-241 (2002).
30. Choi, I.Y., Lee, S.P., Merkle, H. & Shen, J. Single-shot two-echo technique for simultaneous measurement of GABA and creatine in the human brain in vivo. *Magn Reson Med* **51**, 1115-1121 (2004).
31. Du, F., Chu, W.J., Yang, B., Den Hollander, J.A. & Ng, T.C. In vivo GABA detection with improved selectivity and sensitivity by localized double quantum filter technique at 4.1T. *Magn Reson Imaging* **22**, 103-108 (2004).
32. Choi, C. et al. Brain gamma-aminobutyric acid measurement by proton double-quantum filtering with selective J rewinding. *Magn Reson Med* **54**, 272-279 (2005).
33. Badar-Goffer, R.S., Bachelard, H.S. & Morris, P.G. Cerebral metabolism of acetate and glucose studied by 13C-n.m.r. spectroscopy. A technique for investigating metabolic compartmentation in the brain. *Biochem J* **266**, 133-139 (1990).
34. Hong, S., An, L. & Shen, J. Monte Carlo study of metabolite correlations originating from spectral overlap. *J Magn Reson* **341**, 107257 (2022).
35. An, L., Araneta, M.F., Johnson, C. & Shen, J. Effects of carrier frequency mismatch on frequency-selective spectral editing. *Magnetic Resonance Materials in Physics Biology and Medicine* **32**, 237-246 (2019).
36. An, L. et al. Roles of Strong Scalar Couplings in Maximizing Glutamate, Glutamine and Glutathione Pseudo Singlets at 7 Tesla. *Frontiers in Physics* **10** (2022).
37. Feng, Y. et al. Genetic variations in GABA metabolism and epilepsy. *Seizure-Eur J Epilep* **101**, 22-29 (2022).
38. Sarawagi, A., Soni, N.D. & Patel, A.B. Glutamate and GABA Homeostasis and Neurometabolism in Major Depressive Disorder. *Front Psychiatry* **12** (2021).
39. Ben-Menachem, E. et al. The effect of different vigabatrin treatment regimens on CSF biochemistry and seizure control in epileptic patients. *Br J Clin Pharmacol* **27 Suppl 1**, 79S-85S (1989).
40. Riekkinen, P.J., Ylinen, A., Halonen, T., Sivenius, J. & Pitkanen, A. Cerebrospinal fluid GABA and seizure control with vigabatrin. *Br J Clin Pharmacol* **27 Suppl 1**, 87S-94S (1989).
41. Petroff, O.A., Behar, K.L., Mattson, R.H. & Rothman, D.L. Human brain gamma-aminobutyric acid levels and seizure control following initiation of vigabatrin therapy. *J Neurochem* **67**, 2399-2404 (1996).
42. Hasler, G. et al. Normal prefrontal gamma-aminobutyric acid levels in remitted depressed subjects determined by proton magnetic resonance spectroscopy. *Biol Psychiatry* **58**, 969-973 (2005).
43. Schur, R.R. et al. Brain GABA levels across psychiatric disorders: A systematic literature review and meta-analysis of (1) H-MRS studies. *Hum Brain Mapp* **37**, 3337-3352 (2016).
44. Romeo, B., Choucha, W., Fossati, P. & Rotge, J.Y. Meta-analysis of central and peripheral gamma-aminobutyric acid levels in patients with unipolar and bipolar depression. *J Psychiatry Neurosci* **43**, 58-66 (2018).

45. Wang, D., Wang, X., Luo, M.T., Wang, H. & Li, Y.H. Gamma-Aminobutyric Acid Levels in the Anterior Cingulate Cortex of Perimenopausal Women With Depression: A Magnetic Resonance Spectroscopy Study. *Front Neurosci* **13**, 785 (2019).
46. Kantrowitz, J.T. et al. Ventromedial prefrontal cortex/anterior cingulate cortex Glx, glutamate, and GABA levels in medication-free major depressive disorder. *Transl Psychiatry* **11**, 419 (2021).
47. Bottomley, P.A. in Patent, Vol. US4480228A (General Electric Co, US 1984).
48. Murdoch, J.B., Lent, A.H. & Kritzer, M.R. Computer-Optimized Narrow-Band Pulses for Multislice Imaging. *Journal of Magnetic Resonance* **74**, 226-263 (1987).
49. Emsley, L. & Bodenhausen, G. Phase-Shifts Induced by Transient Bloch-Siegert Effects in Nmr. *Chem Phys Lett* **168**, 297-303 (1990).
50. Zhang, Y., An, L. & Shen, J. Fast computation of full density matrix of multispin systems for spatially localized in vivo magnetic resonance spectroscopy. *Medical Physics* **44**, 4169-4178 (2017).
51. Li, S.Z. et al. Determining the Rate of Carbonic Anhydrase Reaction in the Human Brain. *Sci Rep-Uk* **8** (2018).
52. An, L., van der Veen, J.W., Li, S.Z., Thomasson, D.M. & Shen, J. Combination of multichannel single-voxel MRS signals using generalized least squares. *Journal of Magnetic Resonance Imaging* **37**, 1445-1450 (2013).
53. Klose, U. Invivo Proton Spectroscopy in Presence of Eddy Currents. *Magnetic Resonance in Medicine* **14**, 26-30 (1990).
54. de Graaf, R.A. *in vivo* NMR Spectroscopy: Principles and Techniques, Edn. 3. (John Wiley & Sons Ltd, West Sussex, England; 2018).
55. Choi, C.H. et al. Improvement of resolution for brain coupled metabolites by optimized H-1 MRS at 7 T. *Nmr in Biomedicine* **23**, 1044-1052 (2010).
56. Govind, V. H-1-NMR Chemical Shifts and Coupling Constants for Brain Metabolites. *Emagres* **5**, 1347-1362 (2016).

## Acknowledgements

This study (NCT01266577 and NCT00109174) was supported by the Intramural Research Program of the National Institute of Mental Health, National Institutes of Health (IRP-NIMH-NIH, ZIAMH002803).

## Data availability

The data acquired in this study and the code developed to analyze the acquired data are available at https://www.nitrc.org/doi/landing_page.php?doi=10.25790/bml0cm.147

**Author Contributions**

J.S. and L.A. conceived the study and prepared the manuscript. L.A. developed the technology. L.A. and S.H. performed the experiments. M.F.A., T.T., and C.S.J. provided medical support. All the authors read and approved the final manuscript.

**TABLES**

**Table 1** Metabolite ratios (/[tCr]) measured from the ACC of healthy participants (n = 6). The voxel has 60.0% ± 5.2% grey matter, 33.9% ± 3.8% white matter, and 6.1% ± 2.9% cerebrospinal fluid. The within-subject coefficients of variation (CVs) are calculated from the test and re-test measurements of the same voxel.

|      | Metabolite ratio (/[tCr]) | CRLB (%)   | Within-Subject CV (%) |
|------|---------------------------|------------|-----------------------|
| GABA | 0.07 ± 0.01               | 10.0 ± 2.3 | 6.6                   |
| Glu  | 1.25 ± 0.16               | 4.8 ± 1.1  | 5.6                   |
| tCr  | 1                         | 1.1 ± 0.3  | 0                     |
| tCho | 0.29 ± 0.02               | 1.1 ± 0.3  | 0.9                   |

# In Vivo GABA Detection by Single Pulse Editing with One Shot

Li An, Sungtak Hong, Maria Ferraris Araneta, Tara Turon, Christopher S. Johnson, and Jun Shen

Molecular Imaging Branch, National Institute of Mental Health,

National Institutes of Health, Bethesda, MD

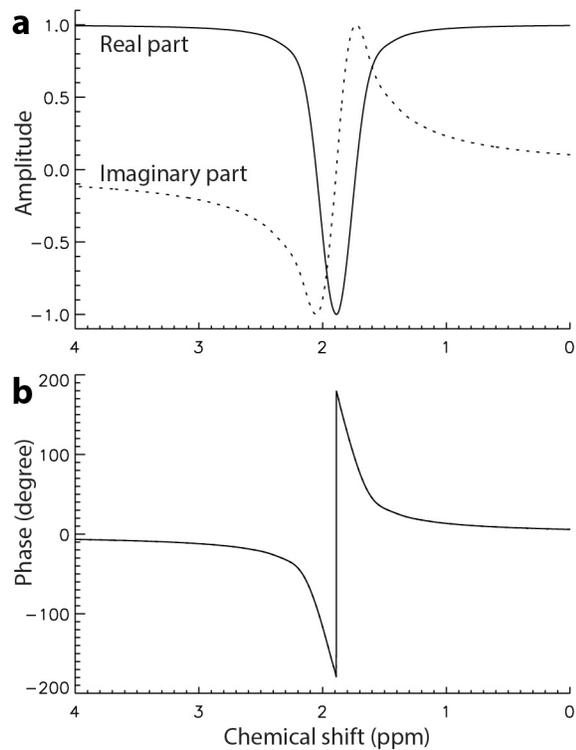

**Fig. S1:** Bloch-Siegert phase shift for SPEOS with a single editing pulse applied at 1.89 ppm. (**a**) Real and imaginary parts of the Bloch-Siegert phasor function. (**b**) Bloch-Siegert phase shift function.

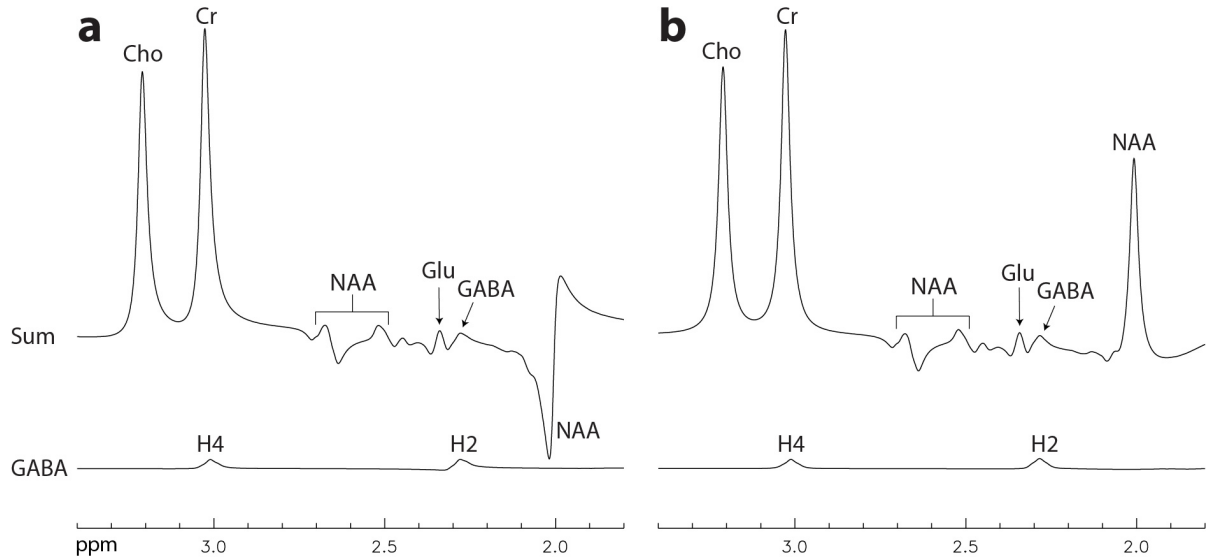

**Fig. S2:** Numerically calculated spectra without (**a**) and with (**b**) Bloch-Siegert phase shift correction. The proposed GABA editing sequence with TE = 76 ms was used in the simulation to calculate the spectra of NAA, GABA, Glu, Cr, and Cho with relative concentrations of 13, 1, 13, 10, and 3, respectively. All spectra were line broadened by 9 Hz using the Lorentzian lineshape.

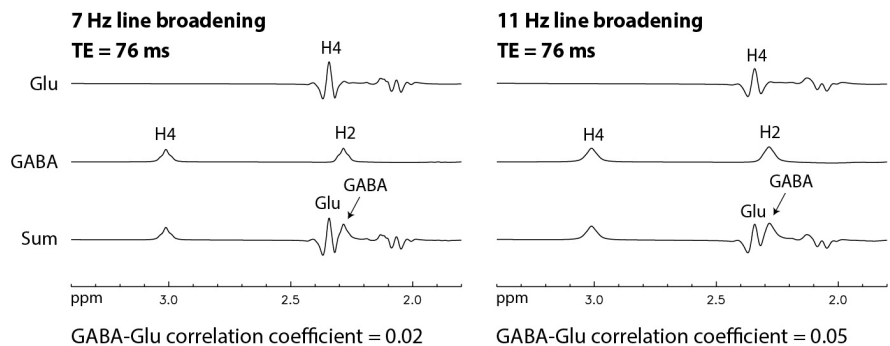

**Fig. S3:** Numerically calculated spectra of Glu, GABA, and their sum at TE = 76 ms with 7 Hz and 11 Hz line broadening.

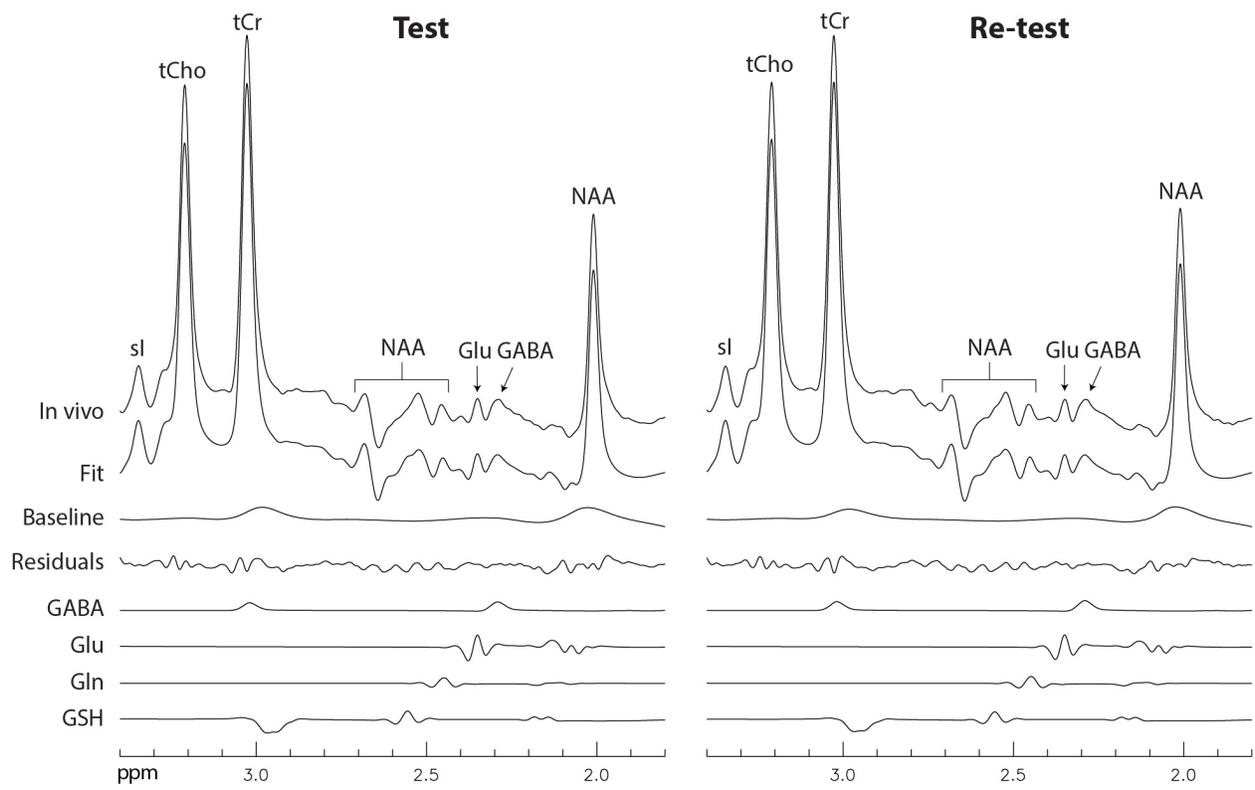

**Fig. S4:** Typical in vivo spectra and fitting details.

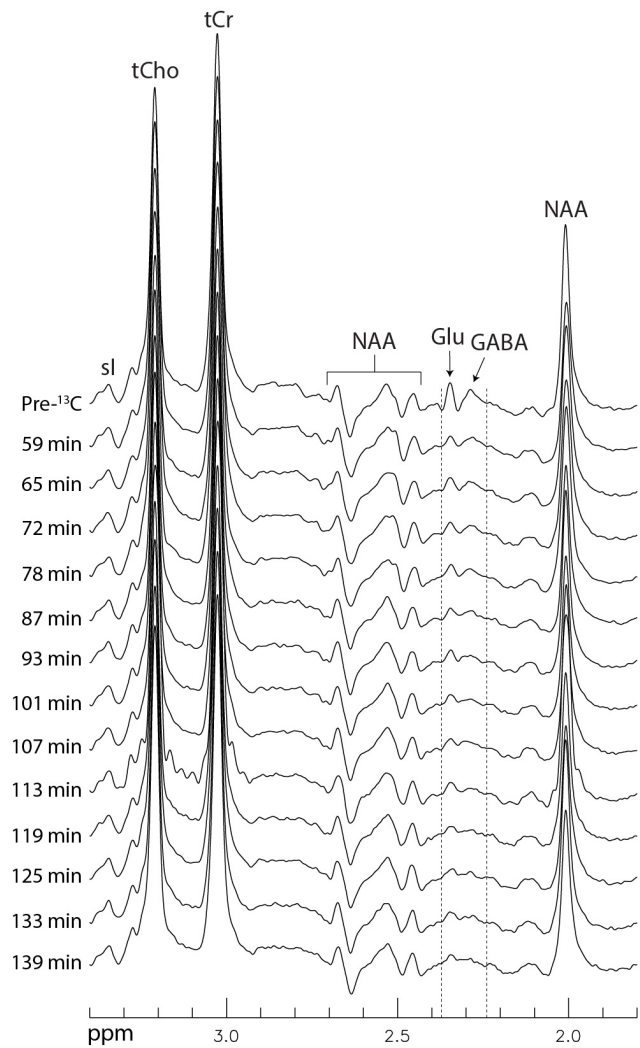

**Fig. S5:** Time-course $^1$H spectra acquired from the ACC of a second healthy participant following oral administration of [U-$^{13}$C]glucose.

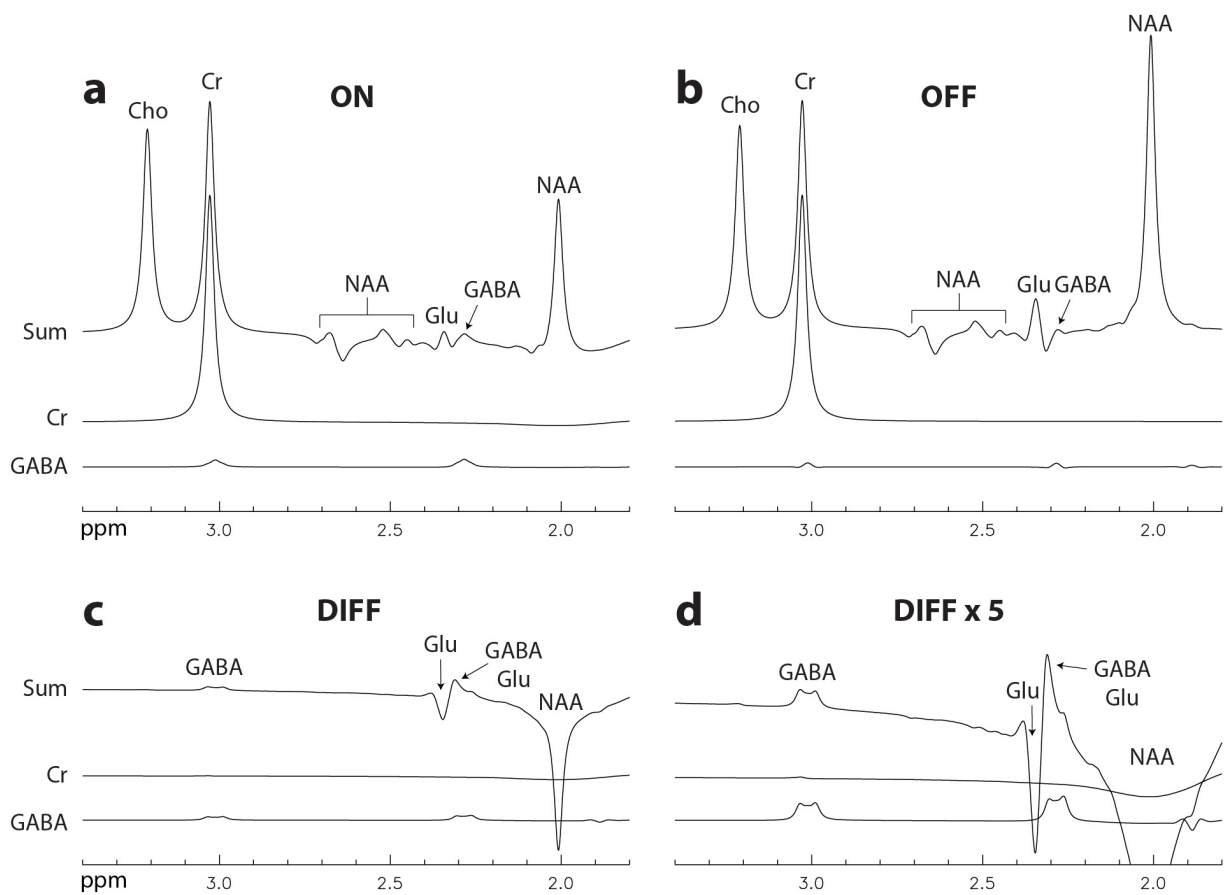

**Fig. S6** Numerically calculated spectra for a two-shot J-difference GABA editing pulse sequence using a single editing pulse. A TE of 76 ms was used to calculate the spectra of NAA, GABA, Glu, Cr, and Cho with relative concentrations of 13, 1, 13, 10, and 3, respectively. All spectra were line broadened by 9 Hz. (**a**) The editing pulse was switched on and applied at 1.89 ppm. (**b**) The editing pulse was switched off. (**c**) Difference spectra. (**d**) The difference spectra were vertically scaled up by a factor of 5.

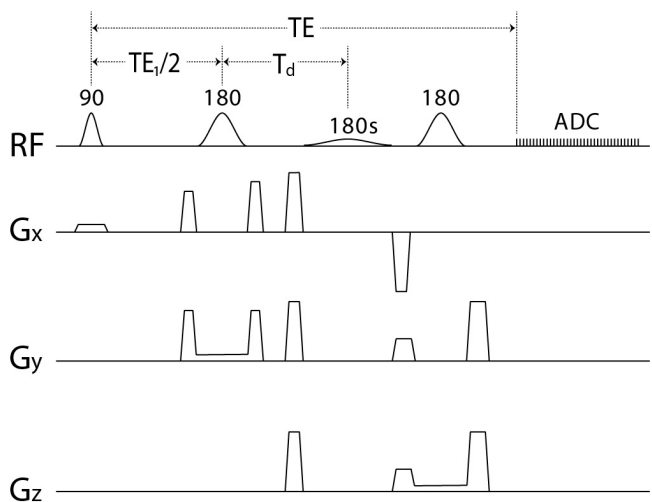

**Fig. S7:** Schematic diagram of the SPEOS pulse sequence for GABA editing. The always-on editing pulse ($180_S$) was applied at 1.89 ppm with a duration of 15 ms. TE = 76 ms; $TE_1$ = 59.1 ms; $T_d$ = 22 ms.

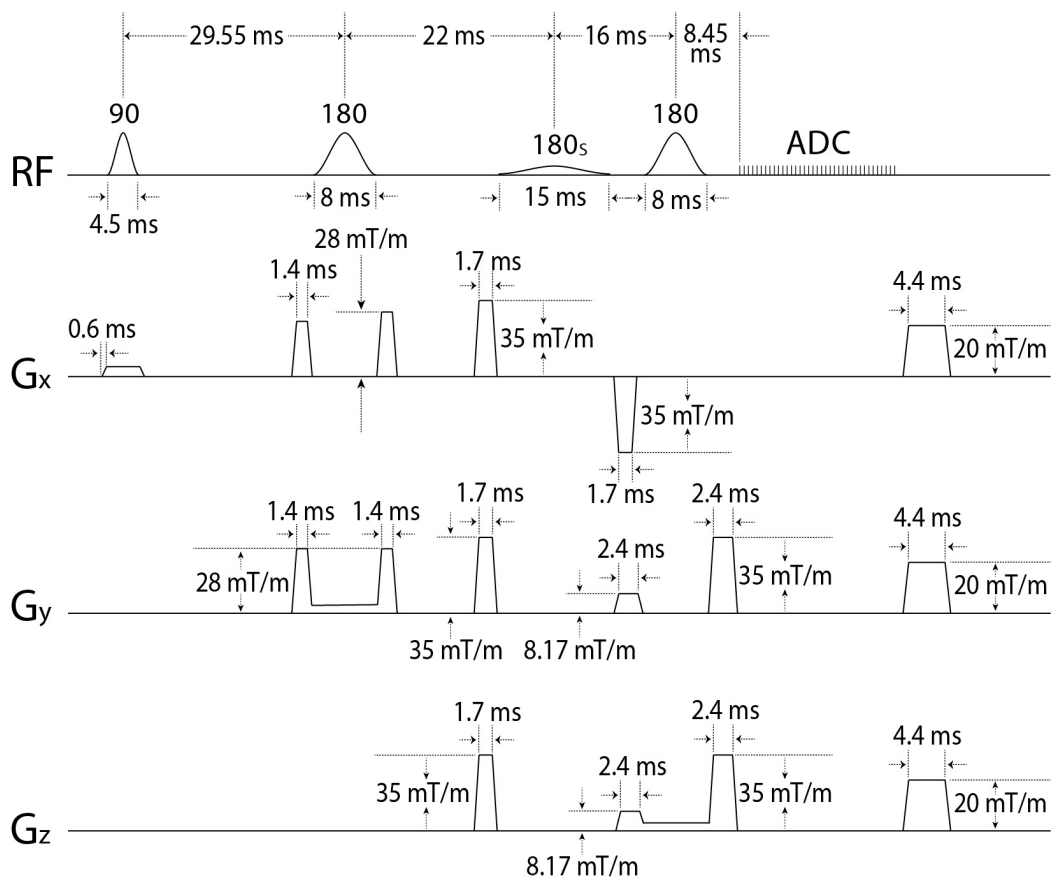

**Fig S8:** Detailed timing diagram of the SPEOS pulse sequence. TR = 2.5 s; Spectral width = 4000 Hz; number of data points = 1024; number of averages = 116; and total scan time = 5 min per spectrum. The slice-selective excitation pulse in the main sequence block was an asymmetric amplitude-modulated pulse with a duration of 4.5 ms, $B_{1,max}$ of 18.6 µT, FWHM bandwidth of 3.1 kHz, and rephase fraction of 0.167. The slice-selective refocusing pulses were amplitude-modulated with a duration of 8.0 ms, $B_{1,max}$ of 18.6 µT, and FWHM bandwidth of 2.0 kHz. The editing pulse ($180_S$) had a duration of 15 ms and $B_{1,max}$ of 1.24 µT, and its amplitude profile was generated by truncating a Gaussian function at one standard deviation on each side. The frequency profile of the editing pulse is shown in Fig. S9.

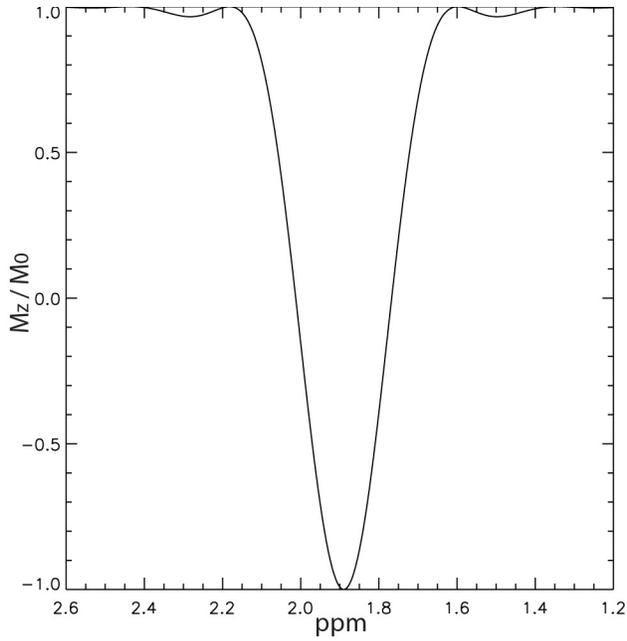

**Fig. S9:** Frequency profile of the editing pulse applied at 1.89 ppm.

## GABA-Glu Correlation

The GABA H2 resonance signal is typically observed within the range of 2.21 – 2.36 ppm at 7 Tesla. The in vivo spectrum in this range can be approximated by a linear combination of the basis spectra of GABA and Glu, described by the following equation:

$$s(n) = c_1 B_1(n) + c_2 B_2(n) + e(n). \tag{S1}$$

Here, $s(n)$ represents the $n^{th}$ data point in the selected spectral range of the in vivo spectrum, $B_1(n)$, $B_2(n)$, $c_1$, and $c_2$ are the basis spectra and concentrations of GABA and Glu in the selected spectral range, respectively. The variable $e(n)$ represents random noise with standard deviation $\sigma$. All variables have real values, as only the real part of the in vivo spectrum is fitted. Equation (S1) can be expressed in matrix form as:

$$\mathbf{s} = \mathbf{Bc} + \mathbf{e} \tag{S2}$$

In this matrix equation, **B** is an N × 2 matrix given by **B** = [**B₁**, **B₂**], where N is the total number of data points in the selected spectral range. **B₁** and **B₂** are column vectors defined as $\mathbf{B_1} = [B_1(1), B_1(2), \ldots, B_1(N)]^T$ and $\mathbf{B_2} = [B_2(1), B_2(2), \ldots, B_2(N)]^T$, respectively, where the superscript T denotes the transpose of the given vector or matrix. Furthermore, **s, c**, and **e** are column vectors defined as $\mathbf{s} = [s(1), s(2), \ldots, s(N)]^T$, $\mathbf{c} = [c_1, c_2]^T$, and $\mathbf{e} = [e(1), e(2), \ldots, e(N)]^T$, respectively. The least square solution to Equation (S2) is given by:

$$\mathbf{c} = (\mathbf{B}^T\mathbf{B})^{-1}\mathbf{B}^T\mathbf{s}. \quad (S3)$$

The covariance matrix of **c** is given by:

$$E[(\mathbf{c} - E(\mathbf{c}))(\mathbf{c} - E(\mathbf{c}))^T] = \sigma^2 \mathbf{F}, \quad (S4)$$

where E represents expectation, and **F** is given by:

$$\mathbf{F} = (\mathbf{B}^T\mathbf{B})^{-1}. \quad (S5)$$

The Pearson's correlation coefficient between GABA and Glu, $r_{12}$, is the off-diagonal element of the 2 × 2 covariance matrix $\sigma^2 \mathbf{F}$ normalized by the square-root of the variances on the main diagonal:

$$r_{12} = F_{12} / \sqrt{F_{11}F_{22}}. \quad (S6)$$

As the inverse of $\mathbf{B}^T\mathbf{B}$, **F** can be readily computed by dividing the adjugate of $\mathbf{B}^T\mathbf{B}$ by its determinant. Therefore, $r_{12}$ can be further expressed as:

$$r_{12} = -\sum B_1(n)B_2(n) / [\sqrt{\sum B_1(n)^2} \sqrt{\sum B_2(n)^2}] = -\mathbf{B_1} \cdot \mathbf{B_2} / (\|\mathbf{B_1}\| \|\mathbf{B_2}\|). \quad (S7)$$

Equation (S7) shows that the correlation coefficient $r_{12}$ is the opposite value of the normalized dot product of **B₁** and **B₂** in the $\mathbf{R}^N$ space. The correlation coefficient is negative if the dot product of the two basis spectra is positive, and vice versa.